# Matter and GPUs: Should the Focus of Our Modeling Classes be Adjusted?

B. J. Fournier & B. M. Wyatt[*]

Tarleton State University, Department of Mathematics, Stephenville, TX 76402, USA

∗e-mail: wyatt@tarleton.edu

## Abstract

We have two basic methods of modeling matter. We can treat matter as a continuum and solve differential equations or we can treat it discretely and solve massive N-body problems. The differential equations produced by meaningful problems can be extremely difficult to formulate and much more difficult to solve, if not impossible. Hence, in most cases, the resultant differential equation is discretized and approximated numerically. We know that matter is discrete, so why all the circular work of taking a discrete phenomenon, putting a continuous model on it, and then discretizing this continuous model into something that is solvable? We do this because the sheer number of particles that make up any meaningful amount of matter is daunting and impossible to handle even with today's largest supercomputer.  One discrete approach to deal with this problem is to group large numbers of particles together and treat them as individual units called quasimolecules. Then pray that the bulk behavior of this model behaves in a similar manner to the matter being studied. The mathematical skills needed to set up such problems are much more attainable than those needed to set up the differential equation, but the N-body problem that ensues usually requires a supercomputer to propagate it through time. Until recently the cost of such machines made this approach out of the financial reach of all except the privileged few. But, thanks to the gaming industry and recent advances in the ease to which one can program modern graphics processing units (GPUs), this is no longer true.  Supercomputing has finally reached the masses. Here we use a simple example of a vibrating string to compare the continuous and discrete approaches to modeling and show how GPUs can enhance an undergraduate modeling experience.

## Introduction

" If, in some cataclysm, all of scientific knowledge were to be destroyed, and only one sentence passed on to the next generations of creatures, what statement would contain the most information in the fewest words? I believe it is the atomic hypothesis (or the atomic fact, or whatever you wish to call it) that all things are made of atoms—little particles that move around in perpetual motion, attracting each other when they are a little distance apart, but repelling upon being squeezed into one another. In that one sentence, you will see, there is an enormous amount of information about the world, if just a little imagination and thinking are applied."  Richard Feynman[1]

Dr. Feynman's statement on the second page of volume one of the Feynman Lectures on Physics is one of the most powerful scientific statements ever made. For most physical processes can be understood if they are thought of in this way. However, modeling each atom in an appreciable amount of matter is far beyond the reach of even today's most powerful computers. Hence, the classical approach to such problems was to treat them as a continuum, an infinitely divisible substance, then model this ideal substance with differential equations. More often than not, the resultant differential equation is too hard to solve exactly so it is discretized and numerically approximated. This approach seems somewhat circular, but until modern computing even this approach was extremely time consuming and most differential equations that could not be solved exactly were not pursued without considerable financial backing. Modern computing of the 1960s allowed researches to numerically attack most differential equations, but solving large n-body problems was still not feasible. In the 80s vector machines came on the scene led by Seymour Cray. These powerful machines were still nowhere close to handling all the particles that actually make up an appreciable amount of matter, but they could handle relatively large n-body problems. This allowed pioneers such as Donald Greenspan to developed Quasimolecular Modeling[2]. This approach attempts to capture the power of the atomic theory but scale it down to a manageable number of bodies. This is done by grouping large aggregates of molecules into quasimolecules with their force interactions and masses adjusted to attain the bulk properties of the material being studied. The main limiting factor to this approach was that these vector machines cost millions of dollars which placed it out of the financial reach of all but the privileged few. Modern computing had put numerical solutions to most differential equations in the hands of the masses but discrete modeling was still not feasible for most. As the scientific community was concentrating on large vector machines for fine grained parallelism and compute clusters for coarse grained parallelism, the gaming industry was focused on creating amazing graphics for video games. Governments may pump large amounts of money into scientific research, but this is only a drop in the bucket compared to the total amount of money gamers put into their gaming computers. Luckily for the scientific community coloring all the pixels of a scene in a video game is embarrassingly parallel. Hence, the graphics processing units (GPUs) that the gaming industry were producing were in reality massive parallel computing machines. The color of a pixel is just a number, so if the GPU could be tricked into producing scientific numbers instead of pixel values you would have a powerful parallel machine for almost no money. The only catch to this idea was that programing GPUs to do scientific work was extremely hard. In 2007 Jen-Hsun Huang and NVIDIA released the compute unified device architecture (CUDA) language that removed this complexity[3]. Supercomputing is now available to the masses.

**Method**

Here we want to compare the discrete method of modeling to the continuous method of modeling. The two aspects we wish to make operant are the differences in mathematical maturity need to apply each method and the qualitative and quantitative differences of each method. To do this we wanted a simple and inexpensive physical event that could be modeled in any classroom. We also wanted a resultant differential equation that one could find a series solution to. This would allow us to directly compare the continuum and discrete modeling approaches. In other words, we did not want to have to discretize the continuum model to approximate its solution. Hence, we chose the vibrating string because it has all the elements we desired and has the added attraction of being the prototype example in most partial differential equation books when introducing hyperbolic equations[4]. Bungee cords, pulleys, and clamps are all readily available at your local hardware store. The only expensive piece of equipment used was a Nikon 1 J3 digital camera which could take videos at 1200 frames per second and cost around 300 dollars. The graphics card used was a GeForce GTX 580 which was placed in an existing machine. The GPU cost around 500 dollars new, but can be bought used for around 150 dollars.

## Physical Experiment

The stiffness, $k$, of the bungee cord was approximated by hanging varied weights to it and determining the distance the cord was stretched relative to the length of the total length of the cord. The stiffness of the cord was nonlinear and inconsistent so an average value for $k$ was used. The cord was clamped on one end, then a weight was hung using a pulley off the other end to set the tension in the cord. The second end of the cord was then clamped. The cord was then pulled laterally to a set distance and released. The resultant action of the cord was filmed with a high speed camera. The distance between the clamps was measured, then the piece of cord between the clamps was cut and weighed to determine the mass per unit length of the cord used. The air resistance, $c$, of the cord was set in the continuum and discrete model so that the oscillations would die off at the same rate as the real bungee cord.

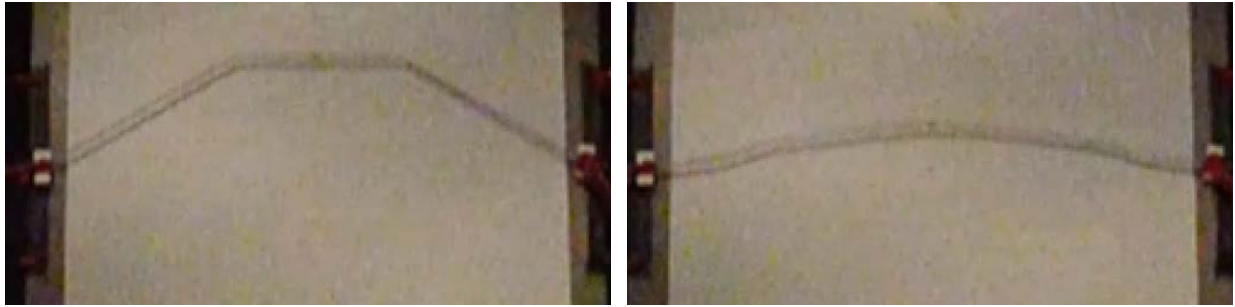

Stills of the actual bungee cord. The first is a snap shot taken at the start of the first oscillation and the second is a snap shot taken at the start of the second oscillation.

Experimental Parameters
Bungee cord: weight - 0.00600 kilograms, Length - 0.67945 meters, stiffness $k$ - 3.9370 Newtons/meter
Resting tension: 4.9050 Newtons
Lateral displacement: 0.18415 meters
Oscillation frequency: 15.002 cycles/second
The measurements presented here are quite crude, but the goal is to present what might be seen in a high school or college freshman modeling class.

## Continuum Solution

To solve the continuum problem, all that is needed to set up the partial differential equation is the tension of the string, $T$, the mass per unit length, $\rho$, of the string, the length of the string, $L$, and the resistance to movement through air, $c$. Then with a couple unrealistic assumptions, that the tension remains constant and that the elemental components of the string only move vertically, you come up with the classic one dimensional damped string equation.

$$U_{tt} = -cU_t + \beta U_{xx}, \quad \beta = \frac{T}{\rho}$$

After about 10 pages of work this equation can be solved by a talented upper level undergraduate mathematics or physics major into the following function:

$$U(x,t) = \sum_{n=1}^{\infty} \left\{ \left[ \frac{cL}{A_n} B_n \sin\left(\frac{A_n}{2L}t\right) + B_n \cos\left(\frac{A_n}{2L}t\right) \right] \sin\left(\frac{n\pi}{L}x\right) \right\}$$

$$A_n = \sqrt{4\beta n^2 \pi^2 - c^2 L^2}$$

$$B_n = \frac{2}{L} \int_0^L f(x) \sin\left(\frac{n\pi}{L}x\right) dx$$

where $f(x)$ is the initial position of the string. If $f(x)$ is relatively simple the student can finally find, with about 5 more pages of work, the actual equation of motion of the string in series form.

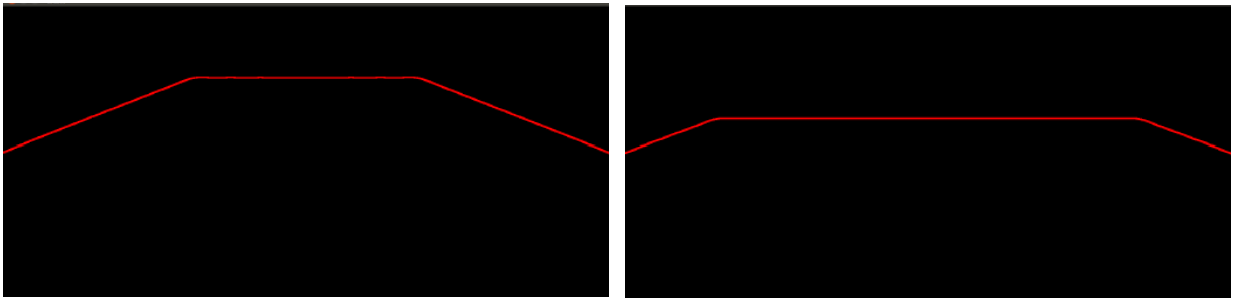

Stills of the continuum model. The first is a snap shot taken at the start of the first oscillation and the second is a snap shot taken at the start of the second oscillation. The frequency given by this model was 17.3 cycles/second. The motion of this model resembled the real string in the early stages but quickly deviated from the true nature of a vibrating string and continued with a stiff, unnatural motion.

### Discrete Solution

To model the motion of the string using the discrete approach we chose the number of quasimolecules we want to use. We divide the total mass of the string by this number to set the mass of the quasimolecules. We then use the tension and the stiffness value $k$ of the cord to set the tension and relative $k$ value of little springs that join the quasimolecules. Note that the $k$ value used to hold the quasimolecules together is a function of length so this value will be the $k$ value of the cord multiplied by the number of quasimolecules plus 1. The force equations are generated, then the system is propagated through time using the leap-frog formulas[5]. The mathematics needed to perform all of these calculations are quite trivial and can be performed at the pre-calculus level. This code is then parallelized so the work can be transferred to the GPU. The CUDA code for the GPU is given below.

```
__global__ void Findforce(float *x, float *y, float *vx, float *vy, float *ax, float *ay, float l, float mass, float k)
{
    int id = threadIdx.x;
    float dx, dy, d2, d, f;

    if(0 < threadIdx.x && threadIdx.x < P-1)
    {
        dx = x[id-1]-x[id];
        dy = y[id-1]-y[id];
        d2 = dx*dx + dy*dy;
        d  = sqrt(d2);
        f = k*(d-l) + Tention;

        ax[id] += (f*dx/d)/mass;
        ay[id] += (f*dy/d)/mass;

        dx = x[id+1]-x[id];
        dy = y[id+1]-y[id];
        d2 = dx*dx + dy*dy;
        d  = sqrt(d2);
        f = k*(d-l) + Tention;
        ax[id] += (f*dx/d)/mass;
        ay[id] += (f*dy/d)/mass;

        ax[id] += (-DAMP*vx[id])/mass;
        ay[id] += (-DAMP*vy[id])/mass;
    }
    __syncthreads();
}
```

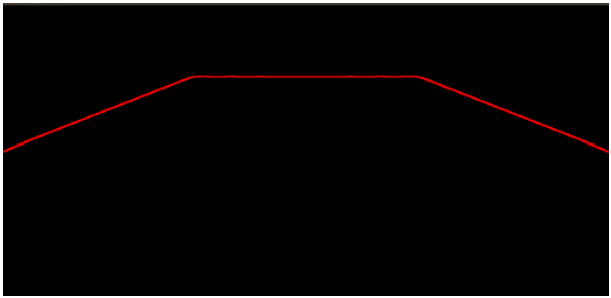 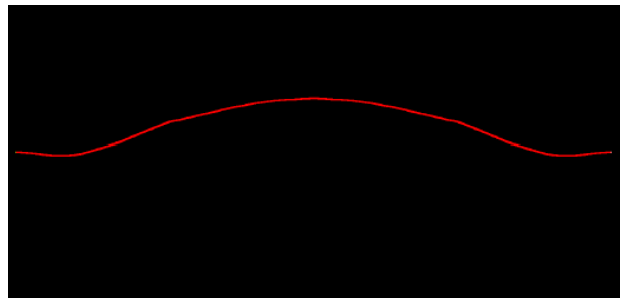

Stills of the continuum model. The first is a snap shot taken at the start of the first oscillation and the second is a snap shot taken at the start of the second oscillation. The frequency given by this model was 19.7 cycles/second. The motion of this model resembled that of the real string throughout the life of its oscillations.

## Conclusion

Both models did a relatively good job at capturing the frequency of the strings vibration, while the discrete model did a much better job at capturing the global nature of the oscillations. The discrete model is much easier to derive, implement, and adapt. The stiffness parameter, $k$, of the bungee cord is nonlinear. Incorporating this into the model would make the PDE unsolvable but would create only a minor adjustment in the discrete model. Hence, the discrete approach is much more robust and adaptable. Discrete modeling is a powerful, robust, and scalable modeling technique that works on a wide range of problems from Astrophysics to nanotechnology. The computational barrier that has kept many from utilizing this modeling approach has all but been removed by the power of GPUs and the CUDA programing language. We don't believe continuum modeling should be removed from modeling courses, but discrete modeling should be added to the curriculum. In addition because the mathematics needed to do discrete modeling is lower, this technique can be used to introduce students to meaningful modeling problems at an earlier age.

```
Appendix A
Complete C and CUDA code

//nvcc particleString.cu -o temp -lglut -lGL -lm
#include <GL/glut.h>
#include <math.h>
#include <stdio.h>

#define PI 3.14159265359

#define X_WINDOW 1000
#define Y_WINDOW 700

#define L 0.67945       //Length of string in Meters
#define D 0.18415       //Laterail displacement of string in Meters
#define MASS  0.006     //Mass of string in Kilograms
#define K 38.6220           //Spring constant of string in Newton/Meters
#define Tention 1.0 //5.3955      //Resting tention in Newtons
#define N 1002          //number of bodies
#define P 1002           //number of bodies per block
#define DAMP 0.00005        //Air resistance

#define X_MAX (L/2.0)
#define X_MIN -(L/2.0)
#define X_SCALE 0.1
```

```c
#define Y_MAX D
#define Y_MIN -D
#define Y_SCALE 0.1

#define TIME_DURATION   1000.0
#define STEP_SIZE       0.0000005
#define TIME_STEP_BETWEEN_VIEWING 100
float *X_CPU, *Y_CPU, *VX_CPU, *VY_CPU, *AX_CPU, *AY_CPU;  //CPU pointers

float *X_GPU, *Y_GPU, *VX_GPU, *VY_GPU, *AX_GPU, *AY_GPU; //GPU pointers

dim3 dimBlock, dimGrid; //Block and Grid Dimensions

void  SetUpCudaDevices() // Sets up the architecture for processes
      {
            //Threads in a block
            dimBlock.x = P;
            dimBlock.y = 1;
            dimBlock.z = 1;

            //Blocks in a grid
            dimGrid.x = 1;
            dimGrid.y = 1;
            dimGrid.z = 1;
      }

void  AllocateMemory()
{
      //Allocate Device (GPU) Memory, & allocates the value of the
specific pointer/array
      cudaMalloc(&X_GPU, N*sizeof(float));
      cudaMalloc(&Y_GPU, N*sizeof(float));
      cudaMalloc(&VX_GPU,N*sizeof(float));
      cudaMalloc(&VY_GPU,N*sizeof(float));
      cudaMalloc(&AX_GPU,N*sizeof(float));
      cudaMalloc(&AY_GPU,N*sizeof(float));

      //Allocate Host (CPU) Memory
      X_CPU  = (float*)malloc(N*sizeof(float)); //(float*) to prevent from
being a void
      Y_CPU  = (float*)malloc(N*sizeof(float));
      VX_CPU = (float*)malloc(N*sizeof(float));
      VY_CPU = (float*)malloc(N*sizeof(float));
      AX_CPU = (float*)malloc(N*sizeof(float));
      AY_CPU = (float*)malloc(N*sizeof(float));
}

float x_machine_to_x_screen(int x)
{
      return( (2.0*x)/X_WINDOW-1.0 );
}
```

```c
float y_machine_to_y_screen(int y)
{
      return( -(2.0*y)/Y_WINDOW+1.0 );
}

/*    Takes machine x and y which start in the upper left corner and go
from zero to X_WINDOW
      left to right and form zero to Y_WINDOW top to bottom and
transslates this into world
      points which are a X_MIN to X_MAX, Y_MIN to Y_MAX window.
*/
float x_machine_to_x_world(int x)
{
      float range;
      range = X_MAX - X_MIN;
      return( (range/X_WINDOW)*x + X_MIN );
}

float y_machine_to_y_world(int y)
{
      float range;
      range = Y_MAX - Y_MIN;
      return(-((range/Y_WINDOW)*y - Y_MAX));
}

/*    Take world  points to screen points
*/
float x_world_to_x_screen(float x)
{
      float range;
      range = X_MAX - X_MIN;
      return( -1.0 + 2.0*(x - X_MIN)/range );
}

float y_world_to_y_screen(float y)
{
      float range;
      range = Y_MAX - Y_MIN;
      return( -1.0 + 2.0*(y - Y_MIN)/range );
}

void  draw_spring(float *x, float *y)
      {
           int i;

           glPointSize(1.0);
           glClear(GL_COLOR_BUFFER_BIT);

           glColor3f(1.0,1.0,0.0);
           glBegin(GL_POINTS);

      glVertex2f(x_world_to_x_screen(x[0]),y_world_to_y_screen(y[0]));
           glEnd();
```

```
            glColor3f(1.0,1.0,1.0);
            glBegin(GL_POINTS);
                glVertex2f(x_world_to_x_screen(x[N-1]),y_world_to_y_screen(y[N-1]));
            glEnd();

            glColor3f(1.0,0.0,0.0);
            for(i = 1; i < N-1; i++)
            {
                glBegin(GL_POINTS);
                glVertex2f(x_world_to_x_screen(x[i]),y_world_to_y_screen(y[i]));
                glEnd();

            }
            glFlush();
        }
__global__ void Findforce(float *x, float *y, float *vx, float *vy, float *ax, float *ay, float l, float mass, float k) // This is the kernel, it is the function that is being fed to the GPU.
{
        int id = threadIdx.x;
        float dx, dy, d2, d, f;

        if(0 < threadIdx.x && threadIdx.x < P-1)
        {
            dx = x[id-1]-x[id];
            dy = y[id-1]-y[id];
            d2 = dx*dx + dy*dy;
            d  = sqrt(d2);
            f = k*(d-l) + Tention;

            ax[id] += (f*dx/d)/mass;
            ay[id] += (f*dy/d)/mass;

            dx = x[id+1]-x[id];
            dy = y[id+1]-y[id];
            d2 = dx*dx + dy*dy;
            d  = sqrt(d2);
            f = k*(d-l) + Tention;
            ax[id] += (f*dx/d)/mass;
            ay[id] += (f*dy/d)/mass;

            ax[id] += (-DAMP*vx[id])/mass;
            ay[id] += (-DAMP*vy[id])/mass;
        }
        __syncthreads();
}

void n_body()
{
        float l, time, dt;
```

```c
	int draw_count,i;
	float particleMass, particleK;
	
	SetUpCudaDevices();
	AllocateMemory();
	
	l = L/(N-1);
	
	time = 0.0;
	dt = STEP_SIZE;
	draw_count = 0;
	
	particleMass = (MASS/L)/(float)N;
	particleK = K*(float(N-1));
	//particleK = K;
	
	X_CPU[0] = -L/2.0;
	Y_CPU[0] = 0.0;
	X_CPU[N-1] = L/2.0;
	Y_CPU[N-1] = 0.0;
	
	for(i=1; i<(N-1); i++)
	{
		X_CPU[i] = -L/2.0 + l*(i);
		if(X_CPU[i] <= 0.0) Y_CPU[i] = (2.0*D/L)*(X_CPU[i]+L/2.0);
		if(X_CPU[i] >  0.0) Y_CPU[i] = D - (2.0*D/L)*(X_CPU[i]);
		VX_CPU[i] = 0.0;
		VY_CPU[i] = 0.0;
	}
	draw_spring(X_CPU, Y_CPU);
	
	printf("\ninter a character in the terminal\n");
	getchar();
	
	while(time < TIME_DURATION)
	{
		for(i=0; i<N; i++)
		{
			AX_CPU[i] = 0.0;
			AY_CPU[i] = 0.0;
		}
		
		//Copy Memory from CPU to GPU
		cudaMemcpyAsync(X_GPU,  X_CPU, N*sizeof(float), cudaMemcpyHostToDevice);
		cudaMemcpyAsync(Y_GPU,  Y_CPU, N*sizeof(float), cudaMemcpyHostToDevice);
		cudaMemcpyAsync(VX_GPU, VX_CPU, N*sizeof(float), cudaMemcpyHostToDevice);
		cudaMemcpyAsync(VY_GPU, VY_CPU, N*sizeof(float), cudaMemcpyHostToDevice);
		cudaMemcpyAsync(AX_GPU, AX_CPU, N*sizeof(float), cudaMemcpyHostToDevice);
```

```
            cudaMemcpyAsync(AY_GPU, AY_CPU, N*sizeof(float), 
cudaMemcpyHostToDevice);
            
            //Launch Kernel
            Findforce<<<dimGrid, dimBlock>>>(X_GPU, Y_GPU, VX_GPU, VY_GPU, 
AX_GPU, AY_GPU, l, particleMass, particleK);
            
            //Copy Memory from GPU to CPU
            cudaMemcpyAsync(AY_CPU, AY_GPU, N*sizeof(float), 
cudaMemcpyDeviceToHost);
        cudaMemcpyAsync(AX_GPU, AX_CPU, N*sizeof(float), 
cudaMemcpyDeviceToHost);
            
            for(i=1; i < (N-1); i++)
            {
                VX_CPU[i] += AX_CPU[i]*dt;
                VY_CPU[i] += AY_CPU[i]*dt;
                X_CPU[i]  += VX_CPU[i]*dt;
                Y_CPU[i]  += VY_CPU[i]*dt;
            }
            
            if(draw_count == TIME_STEP_BETWEEN_VIEWING)
            {
                draw_spring(X_CPU, Y_CPU);
                draw_count = 0;
            }
            
            time = time + dt;
            draw_count++;
        }
}

void display()
{
    //glClear(GL_COLOR_BUFFER_BIT);
    //glFlush();
    
    n_body();
}

int main(int argc, char** argv)
{
    glutInit(&argc,argv);
    glutInitWindowSize(X_WINDOW,Y_WINDOW);
    glutInitWindowPosition(0,0);
    glutCreateWindow("BOX");
    glutDisplayFunc(display);
    glutMainLoop();
}
```

Appendix B:

Complete solution to the damped string equation.

$$U_{tt} + \gamma U_t = \beta U_{xx}$$

Where $\gamma$ and $\beta$ are real numbers greater than zero.

Suppose $U(x,t)$ is separable.

$$U(x,t) = g(t)f(x)$$

$$U_t = g'f, U_{tt} = g''f, \text{ and } U_{xx} = gf''$$

$$g''f + \gamma g'f = \beta gf''$$

$$(g'' + \gamma g')f = (\beta f'')g$$

$$\frac{g'' + \gamma g'}{g} = \frac{\beta f''}{f}$$

Because $g$ is a function of $t$, $f$ is a function of $x$ the only way this could be true is for both $\frac{g''+\gamma g'}{g}$ and $\frac{\beta f''}{f}$ to be a constant. Because $f$ and $g$ are real functions this constant will have to be real.

So we have

$\frac{g''+\gamma g'}{g} = \lambda$ and $\frac{\beta f''}{f} = \lambda$; where $\lambda$ is some real number.

$$\frac{g'' + \gamma g'}{g} = \lambda$$

$$g'' + \gamma g' = \lambda g$$

$$g'' + \gamma g' - \lambda g = 0$$

$$r^2 + \gamma r - \lambda = 0$$

$$r = \frac{-\gamma \pm \sqrt{\gamma^2 + 4\lambda}}{2}$$

1) $g(t) = Ae^{\frac{-\gamma+\sqrt{\gamma^2+4\lambda}}{2}t} + Be^{\frac{-\gamma-\sqrt{\gamma^2+4\lambda}}{2}t}$ if $\gamma^2 + 4\lambda > 0$

2) $g(t) = Ae^{\frac{-\gamma}{2}t} + Bte^{\frac{-\gamma}{2}t}$ if $\gamma^2 + 4 = 0$

3) $g(t) = e^{\frac{-\gamma}{2}t}\left[A \sin\left(\frac{\sqrt{\gamma^2+4\lambda}}{2}t\right) + B \cos\left(\frac{\sqrt{\gamma^2+4\lambda}}{2}t\right)\right]$ if $\gamma^2 + 4 < 0$

$$\frac{\beta f''}{f} = \lambda$$

$$\beta f'' = \lambda f$$

$$\beta f^2 - \lambda f = 0$$

$$\beta r^2 - \lambda = 0$$

$$r^2 = \frac{\lambda}{\beta}$$

$$r = \pm\sqrt{\frac{\lambda}{\beta}}$$

4) $f(x) = Ce^{\sqrt{\frac{\lambda}{\beta}}x} + De^{-\sqrt{\frac{\lambda}{\beta}}x}$ if $\lambda > 0$

5) $f(x) = C = Dx$ if $\lambda = 0$

6) $f(x) = C \sin\left(\sqrt{\frac{-\lambda}{\beta}}x\right) + D \cos\left(\sqrt{\frac{-\lambda}{\beta}}x\right)$ if $\lambda < 0$

1), 2), 3), 4), 5), and 6)

Give 5 different solutions as $\lambda$ ranges over the real numbers

[1] $U(x,t) = \left(Ae^{\frac{-\gamma+\sqrt{\gamma^2+4\lambda}}{2}t} + Be^{\frac{-\gamma-\sqrt{\gamma^2+4\lambda}}{2}t}\right)\left(Ce^{\sqrt{\frac{\lambda}{\beta}}x} + De^{-\sqrt{\frac{\lambda}{\beta}}x}\right)$

$$0 < \lambda$$

[2] $U(x,t) = (A + Be^{-\gamma t})(C + Dx)$

$$\lambda = 0$$

[3] $U(x,t) = \left(Ae^{\frac{-\gamma+\sqrt{\gamma^2+4\lambda}}{2}t} + Be^{\frac{-\gamma-\sqrt{\gamma^2+4\lambda}}{2}t}\right)\left(C \sin\left(\sqrt{\frac{-\lambda}{\beta}}x\right) + D \cos\left(\sqrt{\frac{-\lambda}{\beta}}x\right)\right)$

$$-\frac{\gamma^2}{4} < \lambda < 0$$

[4] $U(x,t) = \left(Ae^{\frac{-\gamma}{2}t} + Bte^{\frac{-\gamma}{2}t}\right)\left(C \sin\left(\sqrt{\frac{-\lambda}{\beta}}x\right) + D \cos\left(\sqrt{\frac{-\lambda}{\beta}}x\right)\right)$

$$\lambda = \frac{-\gamma^2}{4}$$

[5] $U(x,t) = e^{\frac{-\gamma}{2}t}\left[A\sin\left(\frac{\sqrt{\gamma^2+4\lambda}}{2}t\right) + B\cos\left(\frac{\sqrt{\gamma^2+4\lambda}}{2}t\right)\right]\left(C\sin\left(\sqrt{\frac{-\lambda}{\beta}}x\right) + D\cos\left(\sqrt{\frac{-\lambda}{\beta}}x\right)\right)$

$$\lambda < -\frac{\gamma^2}{4}$$

For the first boundary condition $U(0,t) = 0$

- [1] gives $g_1(t) \cdot (C_1 + D_1) = 0$ so $C_1 + D_1 = 0$
- [2] gives $g_2(t) \cdot C_2 = 0$ so $C_2 = 0$
- [3] gives $g_3(t) \cdot D_3 = 0$ so $D_3 = 0$
- [4] gives $g_4(t) \cdot D_4 = 0$ so $D_4 = 0$
- [5] gives $g_5(t) \cdot D_5 = 0$ so $D_5 = 0$

All are still viable solutions

For the second boundary condition $U(L,t) = 0$

- [1] gives $g_1(t) \cdot \left(C_1 e^{\sqrt{\frac{\gamma}{\beta}}L} + D_1 e^{-\sqrt{\frac{\gamma}{\beta}}L}\right) = 0$

    From the two boundary conditions we know that

    $$C_1 + D_1 = 0$$

    $$C_1 e^{\sqrt{\frac{\gamma}{\beta}}L} + D_1 e^{-\sqrt{\frac{\gamma}{\beta}}L} = 0$$

    The above can only be true if $C_1 = D_1 = 0$, which yields the trivial solution.

- [2] gives $g_2(t) \cdot D_2 L = 0$

    From the two boundary conditions we know that
    $$C_2 = 0$$
    $$D_2 = 0$$

    This is a trivial solution.

[3], [4], and [5] all give $g_n(t) \cdot C \sin\left(\sqrt{-\frac{\lambda}{\beta}} L\right) = 0$ with the respective $g(t)$ function.

So $\sqrt{-\frac{\lambda}{\beta}} L = n\pi, n \in \mathbb{I}$

$\sqrt{-\frac{\lambda}{\beta}} = \frac{n\pi}{L}, \lambda = \frac{-\beta n^2 \pi^2}{L^2}$

So $\sqrt{-\frac{\lambda}{\beta}} = \sqrt{\beta \frac{n^2 \pi^2}{L^2 \beta}} = \frac{n\pi}{L}$

and $\sqrt{\gamma^2 + 4\lambda} = \sqrt{\gamma^2 - \frac{4\beta n^2 \pi^2}{L^2}}$ call this $\sqrt{a_n}$

and $\sqrt{-(\gamma^2 + 4\lambda)} = \sqrt{-\gamma^2 + \frac{4\beta n^2 \pi^2}{L^2}}$ call this $\sqrt{-a_n}$

Now throw in super position because we have many solutions for different values of n.

[3] becomes

$$U(x,t) = \sum_{n=?}^{?} \left[A_n e^{\frac{-\gamma + \sqrt{a_n}}{2} t} + B_n e^{\frac{(-\gamma - \sqrt{a_n})}{2} t}\right] \left[C_n \sin\left(\frac{n\pi}{L} x\right)\right]$$

Note: $-\frac{\gamma^2}{4} < \lambda < 0$ so $-\frac{\gamma^2}{4} < -\frac{\beta n^2 \pi^2}{L^2} < 0$

$0 < \frac{\beta n^2 \pi^2}{L^2} < \frac{\gamma^2}{4}$ this may never be true but at most could only be true for a finite number of "n"s.

[4] Most have $-\frac{\gamma^2}{4} = -\frac{\beta n^2 \pi^2}{L^2} => \gamma = 2\frac{\sqrt{\beta} n\pi}{L}$
That is a very restricted drag coefient, but if it exists it gives.

$$U(x,t) = \left[A e^{\frac{-\sqrt{\beta} n\pi}{L} t} + B t e^{\frac{-\sqrt{\beta} n\pi}{L} t}\right] C \sin\left(\frac{n\pi}{L} x\right)$$

[5] becomes

$$U(x,t) = \sum_{n \in \mathbb{I}} e^{-\frac{\gamma}{2} t} \left[A_n \sin\left(\frac{\sqrt{-a_n}}{2} t\right) + B_n \cos\left(\frac{\sqrt{-a_n}}{2} t\right)\right] C_n \sin\left(\frac{n\pi}{L} x\right)$$

Now put in the first initial condition.
$$U(x, 0) = f(x)$$

[3]

$$\sum_{n \text{ Finite}} (A_n + B_n) C_n \sin(\frac{n\pi}{L} x) = f(x)$$

We can not produce f(x) with a finite combinations of " $\sin(\frac{n\pi}{L} x)$"es. Unless f(x) just happens to be this function. Hence [3] is out.

[4] $(A+B)C \sin(\frac{n\pi}{L} x) = f(x)$ this can only work if f(x) is some constant times $\sin(\frac{n\pi}{L} x)$.

Hence [4] is out.

[5]

$$\sum_{n \in \mathbb{I}} B_n \cdot C_n \sin(\frac{n\pi}{L} x) = f(x)$$

This can be done with Fourier work.

Lets clean up [5] by letting $(A_n \cdot C_n) = A_n$ and $(B_n \cdot C_n) = B_n$

so

$$U(x,t) = \sum_{n \in \mathbb{I}} e^{-\frac{\gamma}{2}t} \left[ A_n \sin\left(\frac{\sqrt{-a_n}}{2} t\right) + B_n \cos\left(\frac{\sqrt{-a_n}}{2} t\right) \right] \sin\left(\frac{n\pi}{L} x\right)$$

Now lets clean up [5] even more

$$U(x,t) = \sum_{n \in \mathbb{I}} e^{-\frac{\gamma}{2}t} \left[ A_n \sin\left(\frac{\sqrt{-a_n}}{2} t\right) + B_n \cos\left(\frac{\sqrt{-a_n}}{2} t\right) \right] \sin\left(\frac{n\pi}{L} x\right)$$

Bcause $\sqrt{-a_n} = \sqrt{\gamma^2 + \frac{4\beta n^2 \pi^2}{L^2}}$ then $a_n = a_{-n}$,

$\sin\left(-\frac{n\pi}{L} x\right) = -\sin(\frac{n\pi}{L} x)$, and

$\sin\left(\frac{n\pi}{L} x\right) = 0$ if $n = 0$.

This gives

$$U(x,t) = \sum_{n=1}^{\infty} e^{-\frac{\gamma}{2}t} \left[ (A_{-n} + A_n) \sin\left(\frac{\sqrt{-a_n}}{2} t\right) + (B_{-n} + B_n) \cos\left(\frac{\sqrt{-a_n}}{2} t\right) \right] \sin\left(\frac{n\pi}{L} x\right)$$

Letting $A_{-n} + A_n = A_n$ and $B_{-n} + B_n = B_n$

We have

$$U(x,t) = e^{-\frac{\gamma}{2}t} \sum_{n=1}^{\infty} \left[ A_n \sin\left(\frac{\sqrt{-a_n}}{2} t\right) + B_n \cos\left(\frac{\sqrt{-a_n}}{2} t\right) \right] \sin\left(\frac{n\pi}{L} x\right)$$

This is our lone candidate to satisfy the I.C.

$$U(x,0) = \sum_{n=1}^{\infty} B_n \sin\left(\frac{\sqrt{-a_n}}{2} t\right) = f(x)$$

The $B_n$ can be from Fourier's work as

$$B_n = \frac{2}{L} \int_0^L f(x) \sin\left(\frac{n\pi}{L} x\right) dx$$

This has been done for our starting positions In an attached document.

For the second I.C.; $U_t(x,0) = 0$

$$U(x,t) = e^{-\frac{\gamma}{2}t} \sum_{n=1}^{\infty} \left[A_n \sin\left(\frac{\sqrt{-a_n}}{2} t\right) + B_n \cos\left(\frac{\sqrt{-a_n}}{2} t\right)\right] \sin\left(\frac{n\pi}{L} x\right)$$

$$U_t(x,t) = -\frac{\gamma}{2} U(x,t) + e^{-\frac{\gamma}{2}t} \sum_{n=1}^{\infty} \frac{\sqrt{-a_n}}{2} \left[A_n \cos\left(\frac{\sqrt{-a_n}}{2} t\right) - B_n \sin\left(\frac{\sqrt{-a_n}}{2} t\right)\right] \sin\left(\frac{n\pi}{L} x\right)$$

$$U_t(x,0) = -\frac{\gamma}{2} \sum_{n=1}^{\infty} B_n \sin\left(\frac{n\pi}{L} x\right) + \sum_{n=1}^{\infty} \frac{\sqrt{-a_n}}{2} A_n \sin\left(\frac{n\pi}{L} x\right)$$

$$U_t(x,0) = \left(-\frac{\gamma}{2} \sum_{n=1}^{\infty} B_n + \sum_{n=1}^{\infty} \frac{\sqrt{-a_n}}{2} A_n\right) \sin\left(\frac{n\pi}{L} x\right) = 0$$

So $-\frac{\gamma}{2} B_n + \frac{\sqrt{-a_n}}{2} A_n = 0$

$$A_n = \frac{\gamma}{2} \frac{2}{\sqrt{-a_n}} B_n, \quad A_n = \frac{\gamma}{\sqrt{-a_n}} B_n$$

All together we have with $\sqrt{-a_n} = \frac{\sqrt{4\beta n^2 \pi^2 - \gamma^2 L^2}}{L}$

$$u(x,t) = e^{-\frac{\gamma}{2}t} \sum_{n=1}^{\infty} \left[ \frac{\gamma L}{\sqrt{4\beta n^2 \pi^2 - \gamma^2 L^2}} B_n \sin\left(\frac{\sqrt{4\beta n^2\pi^2 - \gamma^2 L^2}}{2L} t\right) \right.$$
$$\left. + B_n \cos\left(\frac{\sqrt{4\beta n^2\pi^2 - \gamma^2 L^2}}{2L} t\right) \right] \sin\left(\frac{n\pi}{L} x\right)$$

Where $B_n = \frac{2}{L} \int_0^L f(x) \sin\left(\frac{n\pi}{L} x\right) dx$

Now adding in the initial conditions.

$$U(x,t) = \sum_{n=1}^{\infty} \left[ a_n \cos\left(\frac{n\pi\sqrt{a}}{L} t\right) + b_n \sin\left(\frac{n\pi\sqrt{a}}{L} t\right) \right] \sin\left(\frac{n\pi}{L} x\right)$$

$$U(x,0) = \sum_{n=1}^{\infty} a_n \sin\left(\frac{n\pi}{L} x\right) = f(x)$$

$$U_t(x,0) = \sum_{n=1}^{\infty} b_n \left(\frac{n\pi\sqrt{a}}{L}\right) \sin\left(\frac{n\pi}{L} x\right) = g(x) = 0$$

Since $\sin\left(\frac{n\pi}{L} x\right)$ is not zero for all x, then $b_n$ must be zero for all n.

$$f(x) = \sum_{n=1}^{\infty} a_n \sin\left(\frac{n\pi}{L} x\right)$$

$$a_n = \frac{2}{L} \int_0^L f(x) \sin\left(\frac{n\pi}{L} x\right) dx$$

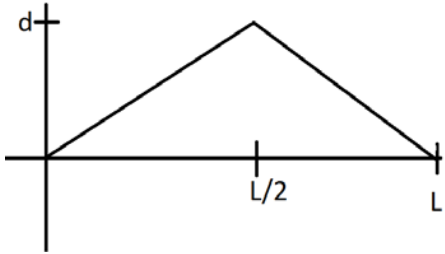

$$f(x) = \begin{cases} \dfrac{2d}{L} x & , 0 \leq x < L/2 \\ -\dfrac{2d}{L} x + 2d & , \dfrac{L}{2} \leq x \leq L \end{cases}$$

$$a_n = \frac{2}{L} \int_0^{\frac{L}{2}} \frac{2d}{L} x \sin\left(\frac{n\pi}{L} x\right) dx + \frac{2}{L} \int_{\frac{L}{2}}^{L} \left(-\frac{2d}{L} x + 2d\right) \sin\left(\frac{n\pi}{L} x\right) dx$$

$$a_n = \frac{4d}{L^2} \int_0^{\frac{L}{2}} x \sin\left(\frac{n\pi}{L} x\right) dx - \frac{4d}{L^2} \int_{\frac{L}{2}}^{L} x \sin\left(\frac{n\pi}{L} x\right) dx + \frac{4d}{L} \int_{\frac{L}{2}}^{L} \sin\left(\frac{n\pi}{L} x\right) dx$$

Let $k = \frac{n\pi}{L}$

Using "u" substitution:

Where

$$u = x, \quad dv = \sin(kx)\, dx$$

$$du = dx, \quad v = -\frac{1}{k}\cos(kx)$$

$$\int x \sin(kx)\, dx = -\frac{x}{k}\cos(kx) + \int \frac{1}{k}\cos(kx)\, dx$$

We find

$$\int x \sin(kx)\, dx = -\frac{x}{k}\cos(kx) + \frac{1}{k^2}\sin(kx) + c$$

Evaluating the first integral:

$$-\frac{4d}{L^2} \cdot \int_0^{\frac{L}{2}} x \sin(kx)\, dx$$

$$-\frac{4d}{L^2} \cdot \left[-\frac{x}{k}\cos(kx) + \frac{1}{k^2}\sin(kx)\right]_0^{\frac{L}{2}}$$

$$-\frac{4d}{L^2} \cdot \left[-\frac{L}{2k}\cos\left(\frac{kL}{2}\right) + \frac{1}{k^2}\sin\left(\frac{kL}{2}\right)\right]$$

Putting $k$ back in.

$$-\frac{4d}{L^2} \cdot \left[-\frac{L}{2}\frac{L}{n\pi}\cos\left(\frac{n\pi}{L} \cdot \frac{L}{2}\right) + \frac{L^2}{n^2\pi^2}\sin\left(\frac{n\pi}{L} \cdot \frac{L}{2}\right)\right]$$

$$-\frac{4d}{L^2} \cdot \left[-\frac{L^2}{2n\pi}\cos\left(\frac{n\pi}{2}\right) + \frac{L^2}{n^2\pi^2}\sin\left(\frac{n\pi}{2}\right)\right]$$

$$-\frac{2d}{n\pi}\cos\left(\frac{n\pi}{2}\right) + \frac{4d}{n^2\pi^2}\sin\left(\frac{n\pi}{2}\right)$$

Evaluating the second integral:

$$-\frac{4d}{L^2} \cdot \int_{\frac{L}{2}}^{L} x \sin(kx)\, dx$$

$$-\frac{4d}{L^2} \cdot \left[-\frac{x}{k}\cos(kx) + \frac{1}{k^2}\sin(kx)\right]_{\frac{L}{2}}^{L}$$

$$-\frac{4d}{L^2} \cdot \left[-\frac{L}{k}\cos(kL) + \frac{1}{k^2}\sin(kL) - \left[-\frac{L}{2k}\cos\left(\frac{kL}{2}\right) + \frac{1}{k^2}\sin\left(\frac{kL}{2}\right)\right]\right]$$

$$-\frac{4d}{L^2} \cdot \left[-\frac{L}{k}\cos(kL) + \frac{1}{k^2}\sin(kL) + \frac{L}{2k}\cos\left(\frac{kL}{2}\right) - \frac{1}{k^2}\sin\left(\frac{kL}{2}\right)\right]$$

Putting $k$ back in.

$$-\frac{4d}{L^2} \cdot \left[-\frac{L^2}{n\pi}\cos(n\pi) + \frac{L^2}{n^2\pi^2}\sin(n\pi) + \frac{L^2}{2n\pi}\cos\left(\frac{n\pi}{2}\right) - \frac{L^2}{n^2\pi^2}\sin\left(\frac{n\pi}{2}\right)\right]$$

$$\frac{4d}{n\pi}\cos(n\pi) - \frac{4d}{n^2\pi^2}\sin(n\pi) - \frac{2d}{n\pi}\cos\left(\frac{n\pi}{2}\right) + \frac{4d}{n^2\pi^2}\sin\left(\frac{n\pi}{2}\right)$$

Evaluating the third integral:

$$\frac{4d}{L} \cdot \int_{\frac{L}{2}}^{L} \sin(kx)\, dx$$

$$\frac{4d}{L} \cdot \left[-\frac{1}{k}\cos(kx)\right]_{\frac{L}{2}}^{L}$$

$$\frac{4d}{L} \cdot \left[-\frac{1}{k}\cos(kL) - \left[-\frac{1}{k}\cos\left(\frac{kL}{2}\right)\right]\right]$$

$$\frac{4d}{L} \cdot \left[-\frac{1}{k}\cos(kL) + \frac{1}{k}\cos\left(\frac{kL}{2}\right)\right]$$

Putting $k$ back in.

$$\frac{4d}{L} \cdot \left[-\frac{L}{n\pi}\cos(n\pi) + \frac{L}{n\pi}\cos\left(\frac{n\pi}{2}\right)\right]$$

$$-\frac{4d}{n\pi}\cos(n\pi) + \frac{4d}{n\pi}\cos\left(\frac{n\pi}{2}\right)$$

Now

$$a_n = \left[-\frac{2d}{n\pi}\cos\left(\frac{n\pi}{2}\right) + \frac{4d}{n^2\pi^2}\sin\left(\frac{n\pi}{2}\right)\frac{4d}{n\pi}\cos(n\pi) - \frac{4d}{n^2\pi^2}\sin(n\pi) - \frac{2d}{n\pi}\cos\left(\frac{n\pi}{2}\right) + \frac{4d}{n^2\pi^2}\sin\left(\frac{n\pi}{2}\right)\right.$$
$$\left. -\frac{4d}{n\pi}\cos(n\pi) + \frac{4d}{n\pi}\cos\left(\frac{n\pi}{2}\right)\right]$$

$$a_n = \cos\left(\frac{n\pi}{2}\right)\left[-\frac{2d}{n\pi} - \frac{2d}{n\pi} + \frac{4d}{n\pi}\right] + \sin\left(\frac{n\pi}{2}\right)\left[\frac{4d}{n^2\pi^2} + \frac{4d}{n^2\pi^2}\right] + \cos(n\pi)\left[\frac{4d}{n\pi} - \frac{4d}{n\pi}\right]$$
$$+ \sin(n\pi)\left[-\frac{4d}{n^2\pi^2}\right]$$

$$a_n = 0 + \sin\left(\frac{n\pi}{2}\right)\left[\frac{8d}{n^2\pi^2}\right] + 0 + 0$$

So

$$a_n = \sin\left(\frac{n\pi}{2}\right)\left[\frac{8d}{n^2\pi^2}\right]$$

Analyzing $a_n = \sin\left(\frac{n\pi}{2}\right)\left[\frac{8d}{n^2\pi^2}\right]$

$$\sin(n\pi/2) = \begin{cases} 0, n = 0 \\ 1, n = 1 \\ 0, n = 2 \\ -1, n = 3 \\ \ldots \end{cases}$$

The only values of n the provide none zero values are the odd ones, so let's change the index.

Let $n = 2k - 1$

Now

$$\sin\left(\frac{(2k-1)\pi}{2}\right) = \begin{cases} 1, k = 0 \\ -1, k = 1 \\ \ldots \end{cases}$$

Which is equivalent to

$$a_n = (-1)^k \left[\frac{8d}{(2k-1)^2\pi^2}\right]$$

Therefore

$$U(x,t) = \sum_{k=1}^{\infty} \frac{8d(-1)^{k+1}}{(2k-1)^2\pi^2} \cos\left(\frac{(2k-1)^2}{L} t\right) \sin\left(\frac{(2k-1)^2\pi}{L} x\right)$$